\documentclass[a4paper]{article}

\usepackage[margin=1in]{geometry} % full-width

\usepackage{amsmath}
\usepackage{amsthm}
\usepackage{amssymb}

\usepackage[utf8]{inputenc}
\usepackage{hyperref}
\hypersetup{
	unicode,
	pdfauthor={Kamel Yamani, Marwa Naïr, Riyadh Baghdadi},
	pdftitle={Automatic Generation of Python Programs Using Context-Free Grammars},
	pdfsubject={Programming Languages},
	pdfkeywords={Automatic Code Generation, Context-Free Grammars (CFG), Backus-Naur Form (BNF), Python},
	pdfproducer={LaTeX},
	pdfcreator={pdflatex}
}

\usepackage[sort&compress,numbers,square]{natbib}
\theoremstyle{plain}

\theoremstyle{definition}

\usepackage{listings}
\usepackage{mdframed}

\usepackage{graphicx, color}
\graphicspath{{fig/}}

\usepackage{algorithm, algpseudocode} % use algorithm and algorithmicx for typesetting algorithms
\usepackage{mathrsfs} % for \mathscr command

\usepackage{lipsum}

\title{\textbf{Automatic Generation of Python Programs Using Context-Free Grammars} }
\author{Kamel Yamani$^{1,2}$\thanks{Both authors contributed equally to this work and share first authorship} \\ \texttt{jm\_yamani@esi.dz}
\and Marwa Naïr$^{1,2*}$ \\ \texttt{jm\_nair@esi.dz} \and Riyadh Baghdadi$^1$ \\ \texttt{baghdadi@nyu.edu}\\\\}

\date{
	$^1$New York University Abu Dhabi (NYUAD), United Arab Emirates \\ 
	$^2$Ecole nationale Supérieure d’Informatique (ESI), Algeria \\ 
}

\begin{document}
	\maketitle
	
	\begin{abstract}
		In recent years, data has emerged as the new gold, serving as a powerful tool for creating intelligent systems. However, procuring high-quality data remains challenging, especially for code. To address this, we developed \textbf{TinyPy Generator}, a tool that generates random Python programs using a context-free grammar. The generated programs are guaranteed to be correct by construction. Our system uses custom production rules (in the Backus-Naur Form (BNF) format) to recursively generate code. This allows us to generate code with different levels of complexity, ranging from code containing only assignments to more complex code containing conditionals and loops. Our proposed tool enables effortless large-scale Python code generation, beneficial for a wide range of applications. \textbf{TinyPy Generator} is particularly useful in the field of machine learning, where it can generate substantial amounts of Python code for training Python language models. Additionally, researchers who are studying programming languages can utilize this tool to create datasets for their experiments, which can help validate the robustness of code interpreters or compilers. Unlike existing research, we have open-sourced our implementation. This allows customization according to user needs and extends potential usage to other languages.
		
	\end{abstract}

	% \tableofcontents
	
	\section{Introduction}
	\label{sec:intro}
        % motivation
        In recent years, obtaining high-quality code data has become increasingly important for powering code intelligence systems, yet several challenges remain. While existing datasets provide large collections of pre-processed real-world code, important context is often missing, making direct execution and correctness verification difficult. Alternatively, scraping public code repositories enables the collection of task-specific data but introduces risks around privacy and intellectual property. Generating code using large language models can help maintain privacy, but the process remains computationally expensive and the resulting programs may not be guaranteed as executable or correct. 

        % our work
        We propose \textbf{TinyPy Generator}, an automatic generator of Python programs using Context-Free Grammars (CFGs) that addresses these limitations. Our method provides complete control over the data distribution while guaranteeing correctness, executability, privacy preservation, and low computational cost. 
        % We focus on Python given its popularity, simplicity, and extensive usage across industry and research. 
        We initially focus on a well-defined subset of the Python language due to its popularity, simplicity, and extensive usage across industry and research. This focus allows for efficient and targeted code generation. However, the underlying framework using CFGs is extensible : with further development, the grammar can be expanded to encompass the full complexity of the Python language.

        % grammar
        Our work builds upon the flexibility of CFGs, which represent a powerful tool for generating code, as they allow for the definition of production rules to recursively generate syntactic structures in a scalable manner. By defining the syntax of Python using CFGs, we ensure that the generated code is not only executable but also well-structured and easily readable. TinyPy Generator uses custom CFG production rules in Backus-Naur form (BNF) to recursively generate code, ensuring that the generated programs are correct by construction. This flexibility allows us to generate code with different levels of complexity, ranging from code containing only assignments to more complex code containing conditionals and loops.     
        
        % generation process
        Our tool's generation process involves several stages, including recursively expanding production rules to generate programs, exception handling, deduplication, and output generation. By handling exceptions and deduplicating code, we ensure that the generated programs are free of errors and redundant code, respectively. Moreover, we execute the generated code and save its output in human-readable manner, making it easier to understand and maintain. Finally, we write both the generated code and it's output (expressed in comment) to a file, allowing for easy integration and usage.
        
        % results
        We demonstrate the efficiency of our tool and diversity of the generated programs through a series of experiments. Our results show that TinyPy Generator is computationally efficient and generates code that is diverse and well-balanced over programming constructs.

        % open-source
        Unlike existing research, we have open-sourced our implementation\footnote{ The source code is available on GitHub at \url{https://github.com/MarwaNair/TinyPy-Generator} } which allows for customization according to user needs and extends potential usage to other programming languages. This makes TinyPy Generator a valuable tool for both the machine learning and programming languages communities.

        % organization of the paper
       In this paper, we detail the design and implementation of TinyPy Generator. First, we provide background on context-free grammars. Next, we explain our grammar to generate Python snippets of varying complexity. We then present the recursive generation process leveraging this grammar. Following that, we showcase results demonstrating efficiency, scalability, and diversity of the generated programs. Finally, we discuss the usefulness of our open-source tool for the machine learning and programming languages communities.

        \section{Background}
	\label{sec:bg}
        \subsection{Context-Free Grammars (CFGs)}
        \label{cfg}
            
        Context-free grammars (CFGs) are a type of formal grammar widely used in computer science and linguistics to describe the syntax of languages. A CFG consists of a set of production rules that specify how strings in the language can be generated from a start symbol through repeated substitution of nonterminal symbols.
        
        Formally, a CFG $G$ is defined by the 4-tuple $G = (V, \Sigma, R, S)$ \cite{sipser_introduction_1997}, where:
        
        \begin{itemize}
            \item $V$ is a finite set of nonterminal symbols that represent syntactic variables,
            \item $\Sigma$ is a finite set of terminal symbols that form the alphabet of the language, such that $\Sigma \cap V = \emptyset$,
            \item $R$ is a finite set of production rules of the form $A \rightarrow \beta$, where $A \in V$ and $\beta \in (V \cup \Sigma)^*$,
            \item $S \in V$ is the start symbol.
        \end{itemize}

        A production rule $A \rightarrow \beta$ means the nonterminal $A$ can be replaced by the string $\beta$, regardless of the context around $A$. This context-free property distinguishes CFGs from other grammars.
        
        Languages generated by CFGs are called context-free languages (CFLs). CFLs have many important properties and can describe a wide range of language constructs, making CFGs fundamental in compiler design, and natural language processing.

        \subsection{Backus-Naur form (BNF)}
        \label{bnf}
        Backus-Naur Form (BNF) is a formal notation used to describe the syntax of programming languages and other formal languages. It provides a way to precisely define the valid sequences of symbols that can be used to construct statements or structures in a given language \cite{mccracken_backus-naur_2003}. A BNF specification consists of a set of derivation rules written in the form:

        \[
        \langle \text{symbol} \rangle ::= \text{expression}
        \]
        
        Where:
        
        \begin{itemize}
            \item $\langle \text{symbol} \rangle$ is a nonterminal symbol enclosed in angle brackets
            \item $::=$ means the symbol on the left must be replaced by the expression on the right
            \item expression consists of sequences of terminals and/or nonterminals separated by vertical bars $|$ to indicate alternatives
        \end{itemize}

\section{Grammar Design}
\label{sec:grammar}

We designed a context-free grammar to automatically generate valid Python code snippets with configurable complexity. The full grammar is provided in \autoref{app:production}. Here we describe the key components that allow us to generate code for assignments and arithmetic expressions, conditionals, loops, and how we combine these to generate programs of increasing complexity.

\subsection{Assignments and Arithmetic Expressions}

We start by defining basic tokens for variables, digits, and arithmetic operators:

\begin{lstlisting}
<variable> ::= a | b | c | ... | z
<digit> ::= 0 | 1 | 2 | ... | 9
<arithmetic_operator> ::= + | - | * |
\end{lstlisting}

Using these tokens, we can generate variable initialization and assignment statements:

\begin{lstlisting}
<initialization> ::= <variable> <equals> <digit>
<simple_assignment> ::= <variable> <equals> <expression>
<advanced_assignment> ::= <variable> <equals> <arithmetic_evaluation>
\end{lstlisting}

\texttt{<initialization>} allows us to declare and initialize variables. \texttt{<simple\_assignment>} assigns the result of an arithmetic \texttt{<expression>} to a variable. \texttt{<advanced\_assignment>} allows more complex arithmetic by assigning the result of a \texttt{<arithmetic\_evaluation>}.

We can build up arithmetic expressions from:

\begin{lstlisting}
<term> ::= <digit> | <variable>
<expression> ::= <term> <arithmetic_operator> <term>
\end{lstlisting}

An \texttt{<expression>} is an arithmetic operation that involves two \texttt{<term>}s separated by an \\ \texttt{<arithmetic\_operator>}. A \texttt{<term>} can be either a \texttt{<digit>} or a \texttt{<variable>}.

Parenthesized sub-expressions are supported through:

\begin{lstlisting}
<enclosed_expression> ::= (<expression>)
\end{lstlisting}

\texttt{<enclosed\_expression>} allows nesting expressions in parentheses.

Complex arithmetic evaluations are constructed from:

\begin{lstlisting}
<arithmetic_evaluation> ::= <arithmetic_evaluation> <arithmetic_operator> <enclosed_expression> | <enclosed_expression>
\end{lstlisting}

\texttt{<arithmetic\_evaluation>} chains together \texttt{<enclosed\_expression>}s with \texttt{<arithmetic\_operator>}s, supporting arbitrarily complex mathematical expressions.

\subsection{Conditionals}

We start by defining basic tokens for relational and logical operators:

\begin{lstlisting}
<relational_operator> ::= < | > | <= | >= | != | ==
<logical_operator_infix> ::= and | or
<logical_operator_prefix> ::= not
\end{lstlisting}

Using these operators, we can generate conditional statements:

\begin{lstlisting}
<simple_if_statement> ::= if <condition> :
<advanced_if_statement> ::= if <chain_condition> :
<simple_elif_statement> ::= elif <condition> :
<advanced_elif_statement> ::= elif <chain_condition> :
<else_statement> ::= else :
\end{lstlisting}

\texttt{<simple\_if\_statement>} and \texttt{<simple\_elif\_statement>} allow basic conditional logic using a single \texttt{<condition>}. \texttt{<advanced\_if\_statement>} and \texttt{<advanced\_elif\_statement>} support chaining multiple conditions with logical operators through \texttt{<chain\_condition>}. \texttt{<else\_statement>} provides an optional else block.

We can build up conditions from:

\begin{lstlisting}
<condition_expression> ::= <expression_identifier> <relational_operator> <expression_identifier> | <expression_identifier> <relational_operator> <digit>
<condition> ::= <optional_not> <condition_expression>
<optional_not> ::= <logical_operator_prefix> <space> | <space>
\end{lstlisting}

A \texttt{<condition\_expression>} compares an \texttt{<expression\_identifier>} with another \\ \texttt{<expression\_identifier>} or a \texttt{<digit>} using a \texttt{<relational\_operator>}. A \texttt{<condition>} can optionally negate the \texttt{<condition\_expression>} using the \texttt{<logical\_operator\_prefix>} \texttt{not}.

Parenthesized sub-conditions are supported through:

\begin{lstlisting}
<enclosed_condition> ::= (<condition>)
\end{lstlisting}

\texttt{<enclosed\_condition>} allows nesting conditions in parentheses.

Complex chained conditions are constructed from:

\begin{lstlisting}
<chain_condition> ::= <chain_condition> <logical_operator_infix> <enclosed_condition> | <logical_operator_prefix> <enclosed_condition> | <enclosed_condition>
\end{lstlisting}

\texttt{<chain\_condition>} combines \texttt{<enclosed\_condition>}s with \texttt{<logical\_operator\_infix>}es or the \texttt{<logical\_operator\_prefix>} \texttt{not}, supporting arbitrarily complex conditional expressions.

\subsection{Loops}

We start by defining basic tokens for loop constructs:

\begin{lstlisting}
<for> ::= for
<in> ::= in
<range> ::= range
\end{lstlisting}

Using these tokens, we can generate for loop statements:

\begin{lstlisting}
<for_header> ::= <for> <expression_identifier> <in> <range> ( <initial> , <final> , <step> ) : | <for> <expression_identifier> <in> <range> ( <initial> , <final> ) :
<for_loop> ::= <for_header> <new_line> <tab_indent> <display>
<advanced_for_loop> ::= <for_loop> | <for_header> <new_line> <tab_indent> <advanced_display>
\end{lstlisting}

\texttt{<for\_header>} defines the loop variable, the range of values to iterate over, and an optional step size. \texttt{<for\_loop>} combines the header with a loop body that displays a value. \texttt{<advanced\_for\_loop>} supports more complex display expressions in the loop body.

We can build up loop components from:

\begin{lstlisting}
<initial> ::= <digit>
<final> ::= <step> * <execution_count> + <initial> - 1
<step> ::= 1 | 2 | 3
<execution_count> ::= 2 | 3
\end{lstlisting}

\texttt{<initial>} specifies the starting value of the range, while \texttt{<final>} calculates the ending value based on the \texttt{<step>} size and the desired \texttt{<execution\_count>} (number of loop iterations).

\subsection{Display}

Displaying values within the generated code snippets is supported through:

\begin{lstlisting}
<display> ::= print( <display_identifier> )
<advanced_display> ::= <display> | print( <display_expression> )
\end{lstlisting}

\texttt{<display>} prints a simple identifier, while \texttt{<advanced\_display>} allows printing more complex expressions involving arithmetic operations.

\subsection{Combined Constructs}

We generate programs of increasing complexity by combining the elements introduced in previous subsections. For this, we have designed rules for six conceptual Levels of Complexity:

\begin{lstlisting}
<level1.1> ::= <initialization> <simple_assignments> <display>
<level1.2> ::= <initialization> <advanced_assignments> <display>
<level2.1> ::= <initialization> <simple_if_statement> <body>
<level2.2> ::= <initialization> <advanced_if_statement> <body>
<level3.1> ::= <initialization> <for_loop>
<level3.2> ::= <initialization> <advanced_assignments> <advanced_for_loop>
<all> ::= <level1.1> | <level1.2> | <level2.1> | <level2.2> | <level3.1> | <level3.2>
\end{lstlisting}

Level 1 consists of variable assignments and has two sublevels: simple assignments with basic arithmetic expressions (\texttt{<level1.1>} ) and advanced assignments with more complex expressions (\texttt{<level1.2>}). Level 2 introduces conditional statements in the form of if-elif-else blocks, with  \texttt{<level2.1>} containing simple conditional logic and \texttt{<level2.2>} adding complexity with arithmetic expressions. Level 3 incorporates looping with for loops - \texttt{<level3.1>} uses a basic for loop while \texttt{<level3.2>} includes advanced arithmetic within the loop body. The start symbol \texttt{<all>} selects one of the six sublevels allowing us to automatically generate Python programs of configurable complexity.

        \section{Generation Process}
        \label{sec:impl}

\begin{figure*}
        \centering
        \includegraphics[scale=0.3]{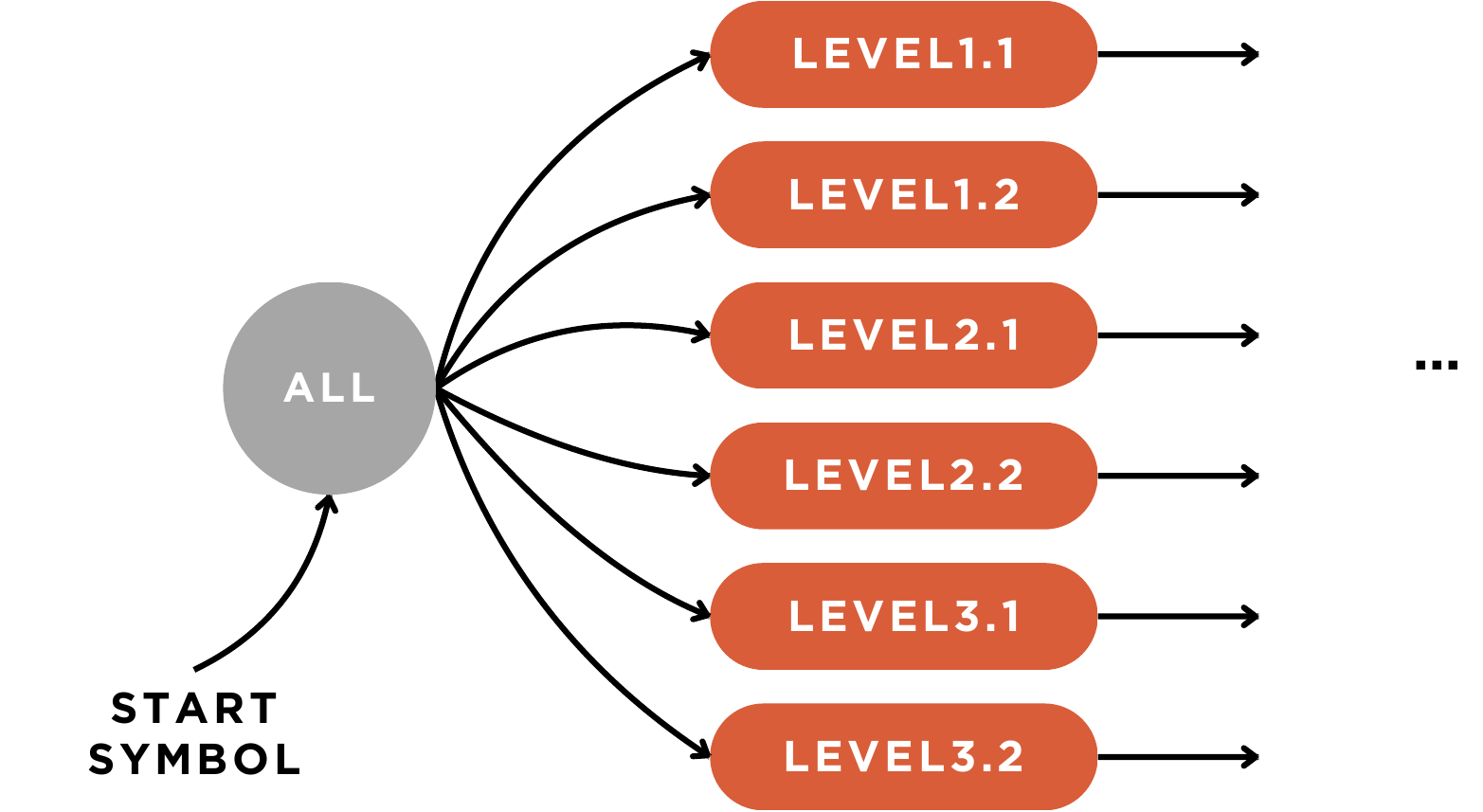}
        \caption{The start symbol "ALL" used in the TinyPy Generator's code generation process.}
        \label{fig:start}
        \end{figure*}

    TinyPy Generator's code generation process is as follows:
    
    \begin{enumerate}
    
    \item \textbf{Start Symbol:} We begin with the root symbol \texttt{<all>} as the starting point of the generation process (refer to \autoref{fig:start}). This symbol represents the entire program and serves as the entry point for the recursive expansion process.
    
    \item \textbf{Level Choice:} A complexity level is randomly selected from the predefined set of levels.
    
    \item \textbf{Recursive Expansion:} We perform a series of recursive expansions using the production rules associated with the chosen complexity level. At each step, a non-terminal symbol is replaced by its corresponding production rule, chosen randomly from the available options for that symbol. This process continues until only terminal symbols remain. We use leftmost derivation because it matches the natural flow of writing code line-by-line from top to bottom.
    
    \item \textbf{Handling Exceptions:} During the expansion process, certain combinations of terminal symbols may lead to exceptions, such as division by zero errors. To ensure the robustness of the generator, these exceptions are caught and handled appropriately.
    
    \item \textbf{Deduplication Check:} To avoid generating duplicate programs, we apply a hash function to the generated code snippet, producing a unique identifier. This hash is compared against the hashes calculated for existing snippets in the corpus. If a match is found, it means the new program is identical to a previously generated one, and the duplicate is discarded, keeping only unique snippets.
    
    \item \textbf{Output Generation:} Once a unique code snippet has been generated, it is executed within a controlled environment, capturing its output. 
    
    \item \textbf{File Writing:} The code snippet, along with its corresponding output, is written to a file for future use. This file serves as the generated corpus of programs.
    
    \end{enumerate}
    
\begin{figure*}
            \centering
            \includegraphics[scale=0.33]{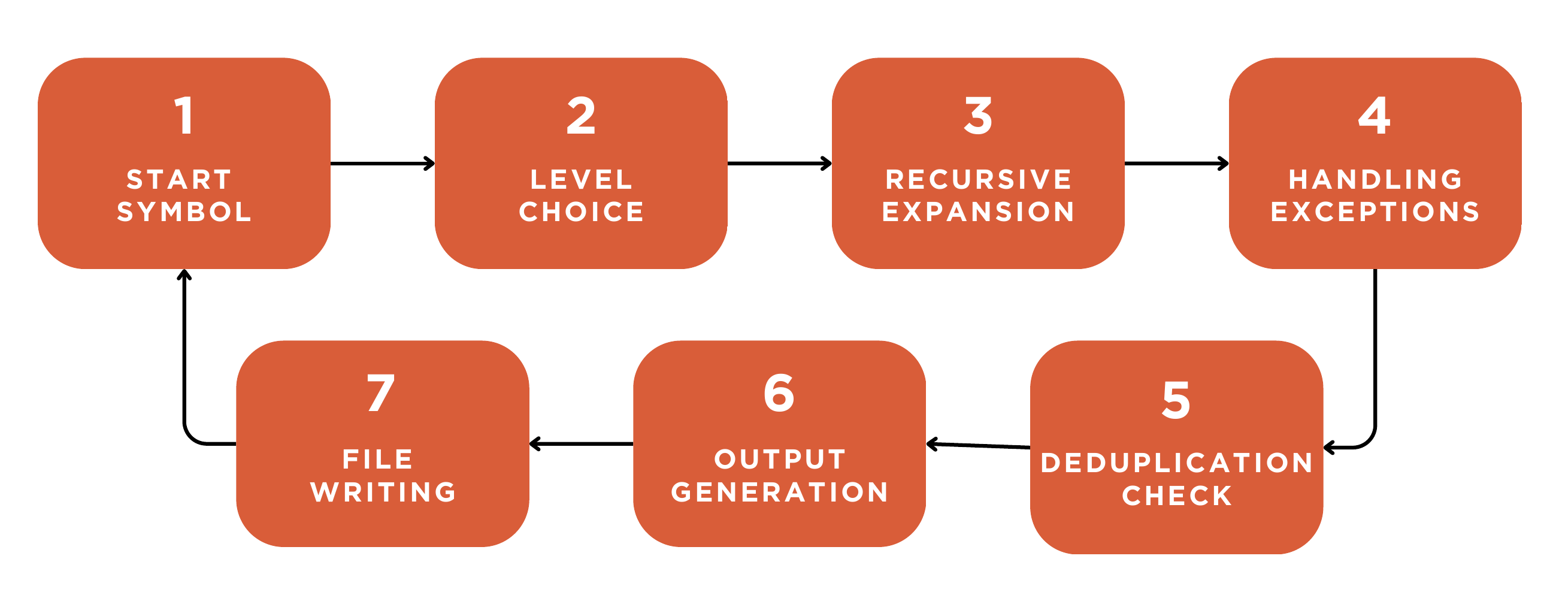}
            \caption{TinyPy Generator's Code Generation Process.}
            \label{fig:process}
        \end{figure*}

    This process repeats iteratively until the desired corpus size is reached, generating a diverse set of programs with varying complexity levels. The entire generation process, illustrated in \autoref{fig:process}, was implemented using Python 3.8.6.
        
        \section{Results}
	\label{sec:res}

        \subsection{Examples}
\begin{figure*}
            \centering
            
            \includegraphics[scale=0.3]{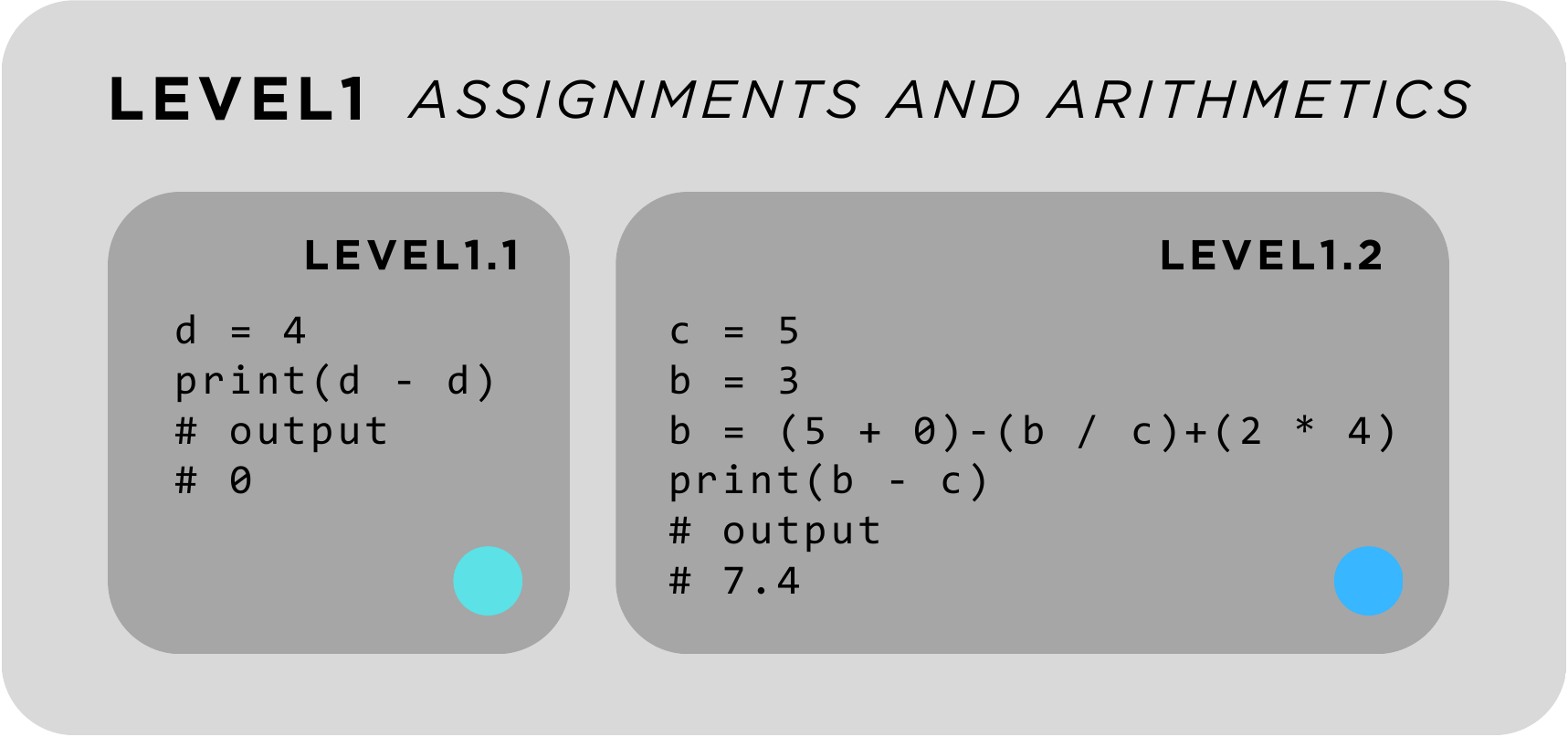}\\
            \vspace{0.4cm}
            \includegraphics[scale=0.3]{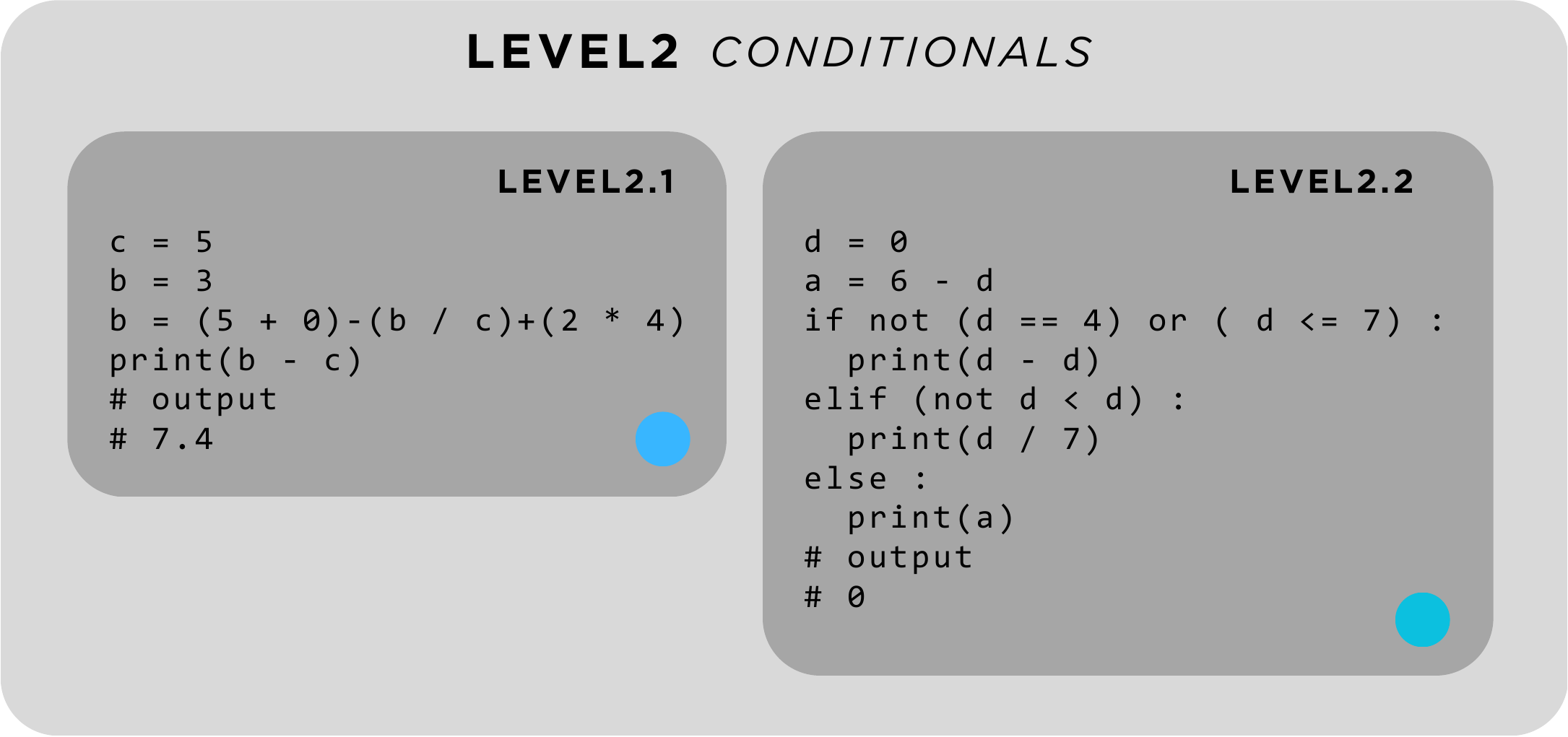}\\
            \vspace{0.4cm}
            \includegraphics[scale=0.3]{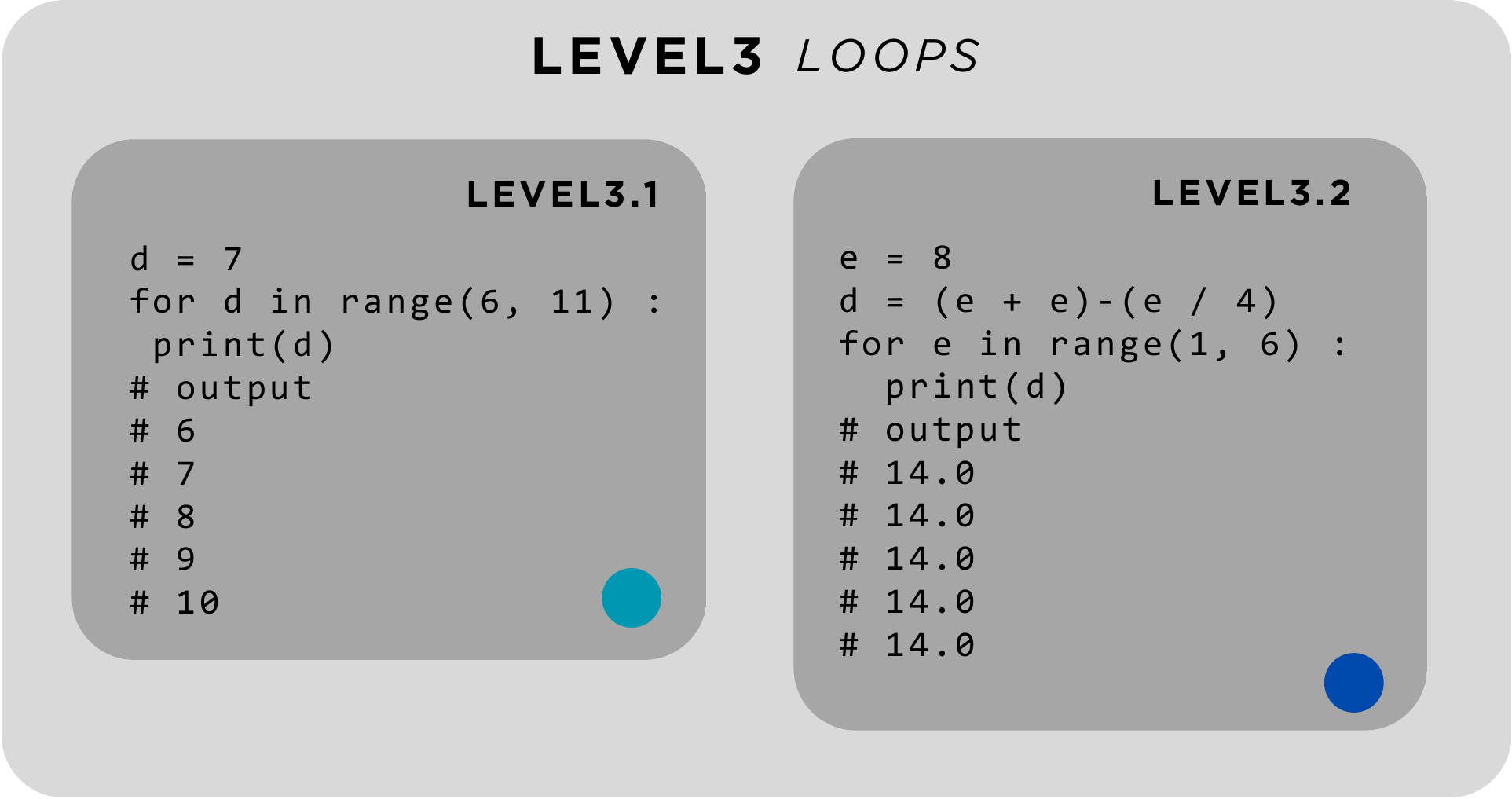}

            \caption{Code Snippets Examples}
            \label{fig:levels}
        \end{figure*}

        In \autoref{fig:levels}, we present examples of generated code snippets across different levels of complexity.

        \subsection{Performance}

        We evaluated the performance of our TinyPy Generator in terms of generation time and memory usage. The tool demonstrated high efficiency, generating up to 1 million unique snippets in approximately 16 minutes using less than 175MB of memory. 

        The \autoref{tab:performance} provides a detailed view of the generation time and memory usage as we increased the number of snippets.
        
        \begin{table}[h]
        \centering
        \begin{tabular}{|c|c|c|}
        \hline
        \textbf{Number of programs} & \textbf{Generation Time} & \textbf{Memory Usage} \\
        \hline
        1,000 & 0.7s & 2.6 MB \\
        10,000 & 7s & 5.2 MB \\
        100,000 & 1m15s & 29.8 MB \\
        1,000,000 & 16m23s & 175.3 MB \\
        \hline
        \end{tabular}
        \caption{Generation time and memory usage for different numbers of programs}
        \label{tab:performance}
        \end{table}

        \subsection{Diversity}
    
        To assess the diversity of the generated code, we computed the frequency distribution of different Python code constructs over three runs of 100,000 samples each, the results are presented in \autoref{tab:code_constructs}
        
        \begin{table}[h]
        \centering
        \begin{tabular}{|c|c|}
        \hline
        \textbf{Code Construct} & \textbf{Average Frequency} \\
        \hline
        Assignments & 0.350 \\
        Conditionals & 0.348 \\
        Loops & 0.302 \\
        \hline
        \end{tabular}
        \caption{Average frequency of different code constructs}
        \label{tab:code_constructs}
        \end{table}

        These results demonstrate the effectiveness of our grammar and generation process in producing a diverse range of programs, covering key constructs such as assignments, conditionals, and loops in similar proportions.

        \section{Applications}
        \label{sec:app}

\begin{figure*}
            \centering
            \includegraphics[scale=0.35]{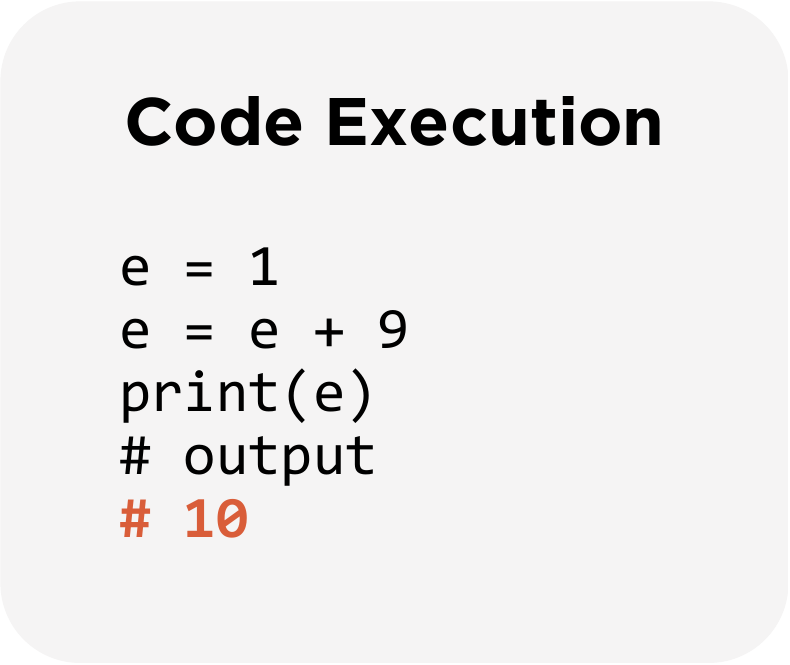}
            \caption{Evaluation on the Code Execution task.}
            \label{fig:eval}
        \end{figure*}

        TinyPy Generator serves as a valuable tool in the fields of machine learning and programming languages research, with the following key applications:
        
        \begin{enumerate}
            \item \textbf{Language Model Training Data Generation}: TinyPy Generator enables the generation of large and diverse Python code datasets, providing valuable training data for Python language models. Its ability to generate syntactically correct code across various topics, from basic constructs to advanced concepts, allows researchers to explore language models' learning capabilities comprehensively.

            As an illustrative example, we trained a 700k parameters NanoGPT \cite{andrej_karpathynanogpt_2022} language model on a dataset of level 1.1 generated by our tool. This resulted in achieving an accuracy of \textbf{96.65\%} in the code execution task (i.e., predicting the output of a given code snippet) presented in \autoref{fig:eval}.
            
            \item \textbf{Validation of Code Interpreters and Compilers}: In programming languages research, TinyPy Generator facilitates the validation of code interpreters or compilers by creating diverse Python code datasets. This allows researchers to thoroughly test the strength and robustness of their tools, identifying potential issues, edge cases, or areas for improvement in handling code with varying levels of complexity and constructs.
        \end{enumerate}

        \begin{figure*}
            \centering
            \includegraphics[scale=0.35]{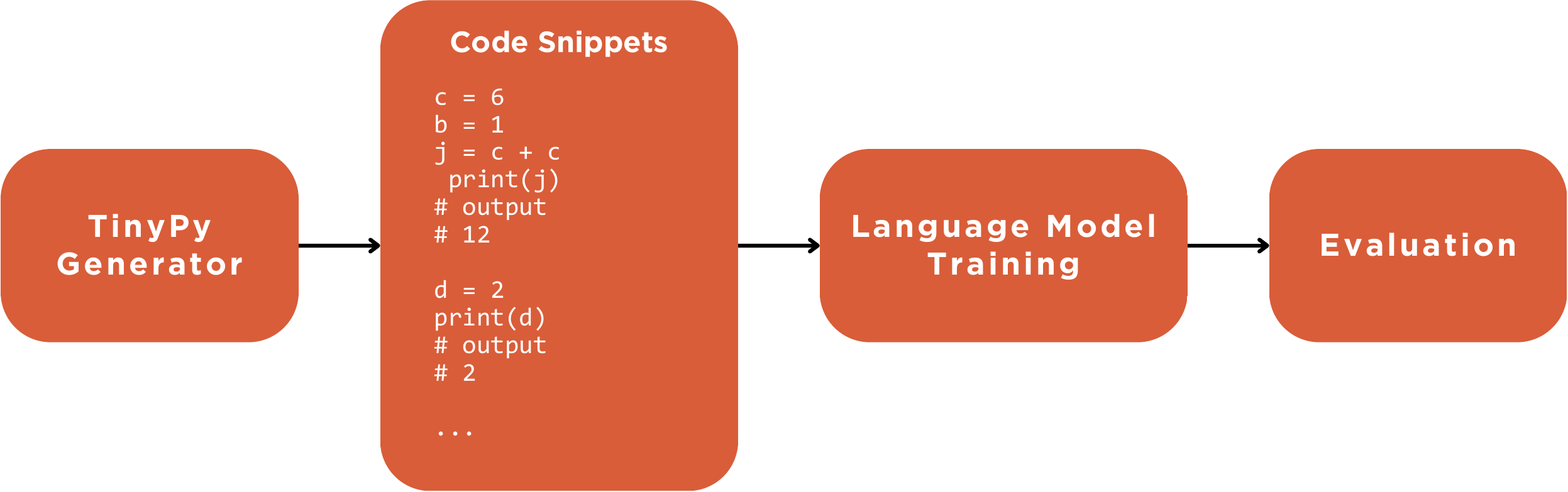}
            \caption{Use case in Machine Learning Research.}
            \label{fig:usecaase}
        \end{figure*}

        \section{Related Works}
	\label{sec:relatedworks}
        Within the scope of our study, a significant contribution is the work presented in \cite{ade-ibijola_syntactic_2018}. This research focused on the development of a code generator designed to assess the programming knowledge of novices. However, the complexity and readability of the generated programs present significant challenges for language modeling. Furthermore, the tool developed in this study is not publicly available, limiting its potential for use or customization in other contexts. This lack of accessibility restricts our ability to build upon this work directly.

        \section{Conclusion}
	\label{sec:ccl}
        We presented TinyPy Generator, a context-free grammar based tool for automatically generating Python code snippets. Our grammar provides fine-grained control over the complexity and composition of the generated programs. The code generator is efficient, scalable, and highly configurable. We demonstrated its usefulness for training machine learning models for Python.
        A key contribution of our work is that we have open-sourced the code generator on GitHub\footnote{ \url{https://github.com/MarwaNair/TinyPy-Generator} }. This allows full customization of the grammar and generation process to suit new applications. Users can tweak the production rules to generate different programming languages beyond Python, expand the subset of Python constructs, control the length and complexity, and more. We provide documentation and usage examples to make it easy for researchers to integrate the tool into their workflows.
        In future work, we plan to expand the capabilities of the tool to generate more complex Python programs with classes, functions, and other constructs.  Overall, we hope our open-source code generator can enable new applications and research at the intersection of machine learning and programming languages.
        
	\bibliography{refs}
	
	\appendix
	
	\section{TinyPy Generator's Complete Set of Production Rules}
	\label{app:production}

        \subsection{Basic Language Constructs}
        This set of rules defines the basic language constructs such as variables, digits, arithmetic operators, relational operators, logical operators, and some special characters and keywords.
        
         \begin{lstlisting}
        <variable> ::= a | b | c | ... | z
        <digit> ::= 0 | 1 | 2 | ... | 9
        
        <arithmetic_operator> ::= + | - | * | /
        <relational_operator> ::= < | > | <= | >= | != | ==
        <logical_operator_infix> ::= and | or
        <logical_operator_prefix> ::= not
        
        <new_line> ::= \n
        <tab_indent> ::= \t
        <bracket_open> ::= (
        <bracket_close> ::= )
        <equals> ::= =
        <colon> ::= :
        <comma> ::= ,
        <if> ::= if
        <elif> ::= elif
        <else> ::= else
        <for> ::= for
        <in> ::= in
        <range> ::= range
        <while> ::= while
        <print> ::= print
        \end{lstlisting}

        \subsection{Expressions and Assignments}
        This set of rules defines how expressions, enclosed expressions, display expressions, and assignments are formed.

        \begin{lstlisting}
        <term> ::= <expression_identifier> | <digit>
        <expression> ::= <term> <space> <operator> <space> <term>
        <enclosed_expression> ::= <bracket_open> <expression> <bracket_close>
        <display_expression> ::= <expression_identifier> <space> <operator> <space> 
        
        <expression_identifier> | <expression_identifier> <space> <operator> <space> <digit>
        <identifier_initialization> ::= <identifier_initialization> <initialization> | <initialization>
        <initialization> ::= <variable> <space> <equals> <space> <digit> <new_line>
        
        <simple_assignments> ::= <variable> <space> <equals> <space> <expression> <new_line> | ""
        <advanced_assignments> ::= <variable> <space> <equals> <space> <simple_arithmetic_evaluation> <new_line> | <variable> <space> <equals> <space> <expression> <new_line> | ""
        <simple_arithmetic_evaluation> ::= <simple_arithmetic_evaluation> <arithmetic_operator> <enclosed_expression> | <enclosed_expression>
        \end{lstlisting}

        \subsection{Conditional Statements}
        This set of rules defines the formation of simple and advanced conditional statements (if, elif, else).

        \begin{lstlisting}
        <simple_if_statement> ::= <if> <space> <condition> <space> <colon> <new_line>
        <advanced_if_statement> ::= <if> <space> <chain_condition> <space> <colon> <new_line>
        
        <simple_elif_statement> ::= <elif> <space> <condition> <space> <colon> <new_line>
        <advanced_elif_statement> ::= <elif> <space> <chain_condition> <space> <colon> <new_line>
        
        <else_statement> ::= <else> <space> <colon> <new_line>
        
        <chain_condition> ::= <chain_condition> <space> <logical_operator_infix> <space> <enclosed_condition> | <logical_operator_prefix> <space> <enclosed_condition> | <enclosed_condition>
        <enclosed_condition> ::= <bracket_open> <condition> <bracket_close>
        <condition> ::= <optional_not> <condition_expression> | <condition_expression>
        <condition_expression> ::= <expression_identifier> <space> <relational_operator> <space> <expression_identifier> | <expression_identifier> <space> <relational_operator> <space> <digit>
        <optional_not> ::= <logical_operator_prefix> <space> | <space>
        \end{lstlisting}

        \subsection{Loop Constructs}
        This set of rules defines the formation of for loop headers and loops.

        \begin{lstlisting}
        <for_header> ::= <for> <space> <expression_identifier> <space> <in> <space> <range> <bracket_open> <initial> <comma> <space> <final> <comma> <space> <step> <bracket_close> <space> <colon> | <for> <space> <expression_identifier> <space> <in> <space> <range> <bracket_open> <initial> <comma> <space> <final> <bracket_close> <space> <colon>
        
        <initial> ::= <digit>
        <final> ::= <step> * <execution_count> + <initial> - 1
        <step> ::= 1 | 2 | 3
        <execution_count> ::= 2 | 3
        
        <for_loop> ::= <for_header> <new_line> <tab_indent> <display>
        <advanced_for_loop> ::= <for_loop> | <for_header> <new_line> <tab_indent> <advanced_display>
        \end{lstlisting}

        \subsection{Display and Levels}
        This set of rules defines how print statements are formed and how different levels of language constructs are combined.

        \begin{lstlisting} 
        <display> ::= <print> <bracket_open> <display_identifier> <bracket_close>
        <advanced_display> ::= <display> | <print> <bracket_open> <display_expression> <bracket_close>
        
        <level1.1> ::= <identifier_initialization> <simple_assignments> <advanced_display>
        <level1.2> ::= <identifier_initialization> <advanced_assignments> <advanced_display>
        <level2.1> ::= <identifier_initialization> <simple_if_statement> <tab_indent> <display> | <identifier_initialization> <simple_if_statement> <tab_indent> <display> <new_line> <simple_elif_statement> <tab_indent> <display> <new_line> <else_statement> <tab_indent> <display> | <identifier_initialization> <simple_if_statement> <tab_indent> <display> <new_line> <else_statement> <tab_indent> <display>
        <level2.2> ::= <identifier_initialization> <advanced_assignments> <advanced_if_statement> <tab_indent> <advanced_display> | <identifier_initialization> <advanced_assignments> <advanced_if_statement> <tab_indent> <advanced_display> <new_line> <advanced_elif_statement> <tab_indent> <advanced_display> <new_line> <else_statement> <tab_indent> <advanced_display> | <identifier_initialization> <advanced_assignments> <advanced_if_statement> <tab_indent> <advanced_display> <new_line> <else_statement> <tab_indent> <advanced_display>
        <level3.1> ::= <identifier_initialization> <for_loop>
        <level3.2> ::= <identifier_initialization> <advanced_assignments> <advanced_for_loop>
        
        <all> ::= <level1.1> | <level1.2> | <level2.1> | <level2.2> | <level3.1> | <level3.2>
        \end{lstlisting}

\end{document}